\documentclass{article}
\usepackage{graphicx}
\usepackage{amsmath}
\usepackage{natbib}
\usepackage{hyperref}
\usepackage{comment}
\usepackage{array}
\usepackage{tikz}

\begin{document}

\title{Target-Date Funds: A State-of-the-Art Review with Policy Applications to Chile's Pension Reform}

\author{Fernando Suárez\thanks{Corresponding author: research@fintual.com} \\
\textit{Fintual AGF, Chile.}
\and
José Manuel Peña \\
\textit{Fintual AGF, Chile.} \\
\and
Omar Larré \\
\textit{Fintual AGF, Chile}
}

\date{\today}
\maketitle

\begin{abstract}
This review paper explores the evolution and implementation of target-date funds (TDFs), specifically focusing on their application within the context of Chile's 2025 pension reform. The introduction of TDFs marks a significant shift in Chile's pension system, which has traditionally relied on a multifund structure (essentially a target-risk funds system). We offer a comprehensive review of the theoretical foundations and practical considerations of TDFs, highlighting key challenges and opportunities for Chilean regulators and fund managers. Notably, we recommend that the glide path design should be dynamic, incorporating adjustments based on total accumulated wealth, with particular flexibility depending on each investor's risk tolerance. Furthermore, we propose that the new benchmark for generational funds should feature a wide deviation band relative to the new benchmark portfolio, which could foster a market with more investment strategies and better competition among fund managers, encourage the inclusion of alternative assets, and foster greater diversification. Lastly, we highlight the need for future work to define a glide path model that incorporates the theoretical frameworks described, tailored to the unique parameters of the Chilean pension system. These recommendations aim to optimize the long-term retirement outcomes for Chilean workers under the new pension structure.

\end{abstract}

\section{Introduction}

In 1994, Donald Luskin and Larry Tint of Wells Fargo Investment Advisors \citep{patent} introduced the first target-date funds (TDFs) to the asset management industry. These funds, though straightforward, represent a smart approach to long-term investing. The concept is simple: investors commit their capital until a predetermined target date, during which the asset allocation shifts dynamically over time to adjust for the remaining investment horizon. This gradual adjustment in portfolio composition, widely known as the glide path, typically begins with a heavy allocation to riskier assets such as stocks, and then gradually transitions to a portfolio that is more concentrated in less risky instruments, such as bonds \citep{estrada1}, as the investor approaches retirement.

Although TDFs have grown immensely in popularity, they are just one example of a broader approach to investing known as life cycle investing. This approach adjusts an individual's asset allocation as they progress through different life stages, taking into account factors such as age, financial goals, and risk tolerance. Life cycle investments are thus designed to ensure that the portfolio remains aligned with the investor's changing needs. The theoretical foundation for these strategies, which predates their modern application, can be traced back to \cite{samuelson1} and \cite{MERTON1}, who extended the static portfolio theory of \cite{markowitz} by introducing dynamic programming models and calculus of variations techniques. This innovation enabled the determination of an optimal, evolving asset allocation over an investor’s lifetime, accounting for path-dependent investment and consumption decisions.

Despite their solid theoretical foundation, life cycle funds only began gaining significant traction after the Pension Protection Act of 2006, which designated them as one of the Qualified Default Investment Alternatives (QDIAs) in U.S. employer-sponsored retirement plans. QDIAs are investment options that plan sponsors can use as default choices for participants who do not actively select an investment, offering a safe harbor under ERISA guidelines \citep{pensionact}. Since then, the market for these funds has expanded dramatically, with a greater variety of offerings, increasing assets under management, and declining expense ratios \citep{lessons, elton}.

With this international context established, we turn to Chile. This year, after years of debate, Chile’s Congress passed a comprehensive pension reform \citep{reforma}. Among other changes, the reform introduces target-date funds—referred to as “fondos generacionales”—as the default investment option. Although Chile is not the first Latin American country to incorporate TDFs into its pension system—Mexico adopted them in 2019 \citep{consar}—the recent reform leaves key details unresolved.

For instance, the specific glide path strategy has yet to be defined, and the responsibility for establishing it falls to the regulator, the Superintendencia de Pensiones. In addition to defining the benchmark glide path, the regulator must also determine the maximum allowable deviations from that benchmark, as well as certain investment limits aimed at mitigating specific portfolio risks. Consequently, both regulators and fund managers now face the challenge of determining the most suitable investment policy for the future target-date pension fund portfolios.

Against this backdrop, the aim of this paper is to provide a solid bibliographical foundation and a state-of-the-art overview of life cycle investing and TDFs. By examining lessons learned from these strategies, we hope to offer valuable insights to fund managers and Chilean regulators as they design effective glide path frameworks.

The paper is organized into two sections. The first provides a bibliographical review, covering foundational models through to the current state of the target-date fund industry. The second examines Chile’s pension system and investment framework, concluding with a discussion of the critical factors that regulators and fund managers must consider when developing glide path strategies.

\section{Literature Review}
\subsection{Foundations of Life Cycle Investment Theory}
The theoretical foundations of life cycle investment strategies can be traced back to \cite{samuelson1}, who extended Markowitz’s classical portfolio selection theory in 1969 to incorporate multiple time periods. Markowitz’s static framework, which seeks to select the optimal set of assets to maximize returns under risk constraints (often represented by volatility), was adapted by Samuelson to address how optimal portfolio decisions evolve over an individual’s life.

Samuelson framed the problem using the Ramsey model \citep{ramsey}, which is based on maximizing intertemporal utility through an optimal consumption function. In this model, individuals face decisions about how much to consume in each period and how to allocate their wealth between risky and risk-free assets. Samuelson's life cycle model was formulated as an optimal control problem, solvable using the calculus of variations, with the objective of maximizing the expected utility of consumption over time.

More formally, let us define \( W_t \) as total wealth at time \( t \), \( C_t \) as the consumption rate at time \( t \), and \( w_t \) as the proportion of wealth invested in risky assets at time \( t \). Wealth evolves according to the constraint
\[
C_t = W_t - W_{t+1} \left[ (1 - w_t)(1 + r) + w_t Z_t \right],
\]
where \( r \) is the return on the risk-free asset and \( Z_t \) is the return on the risky asset at time \( t \). Samuelson looked for an optimal policy $J$ such that

\[
J = \max_{\{C_t, w_t\}} E \left[ \sum_{t=0}^{T} \frac{U(C_t)}{(1+\rho)^{t}}  \right],
\]
where \( U(\cdot) \) is the utility function and \( \rho \) is the discount rate. The model assumes no bequest, that is, \( W_{T+1} = 0 \), and the proportion of wealth allocated to risky assets, \( w_t \), is limited to lie between 0 and 1.

Samuelson’s work was extended by his student, Robert Merton, who introduced a continuous-time version \citep{MERTON1} and examined portfolio choices under two types of utility functions: Constant Relative Risk Aversion (CRRA) and Constant Absolute Risk Aversion (CARA). In the CRRA case, the portfolio allocation remains constant over time, whereas under CARA, the portfolio exhibits a descending glide path, with a decreasing allocation to risky assets as the investor ages. Merton also introduced a “hump-shaped” savings curve, where wealth is accumulated early in life and dissaved closer to retirement.

Later studies expanded the framework by integrating \textit{human capital} (defined as the value of an individual's labor income) into the decision-making process. \cite{merton2} analyzed the impact of labor flexibility on investment decisions, finding that individuals with more flexible labor supply tend to take on greater investment risk early in their careers, as this flexibility offers insurance against poor investment outcomes. \cite{gomes} extended the framework to include consumption and asset accumulation decisions, showing that younger individuals typically prefer equities, with their allocation to risky assets declining over time. However, they also found that greater labor flexibility leads to higher equity holdings before retirement. Their work emphasized the welfare costs of restricting individuals to default investment options in defined contribution plans, particularly when these options do not align with investors' preferences.

\cite{cocco} further incorporated labor income into life cycle models, showing that higher income—particularly when positively correlated with stock returns—leads to greater equity investments. Cocco’s model also revealed that younger individuals consume most of their income, while retirement savings gradually increase with age.

\begin{figure}
    \centering
    \includegraphics[width=0.9\linewidth]{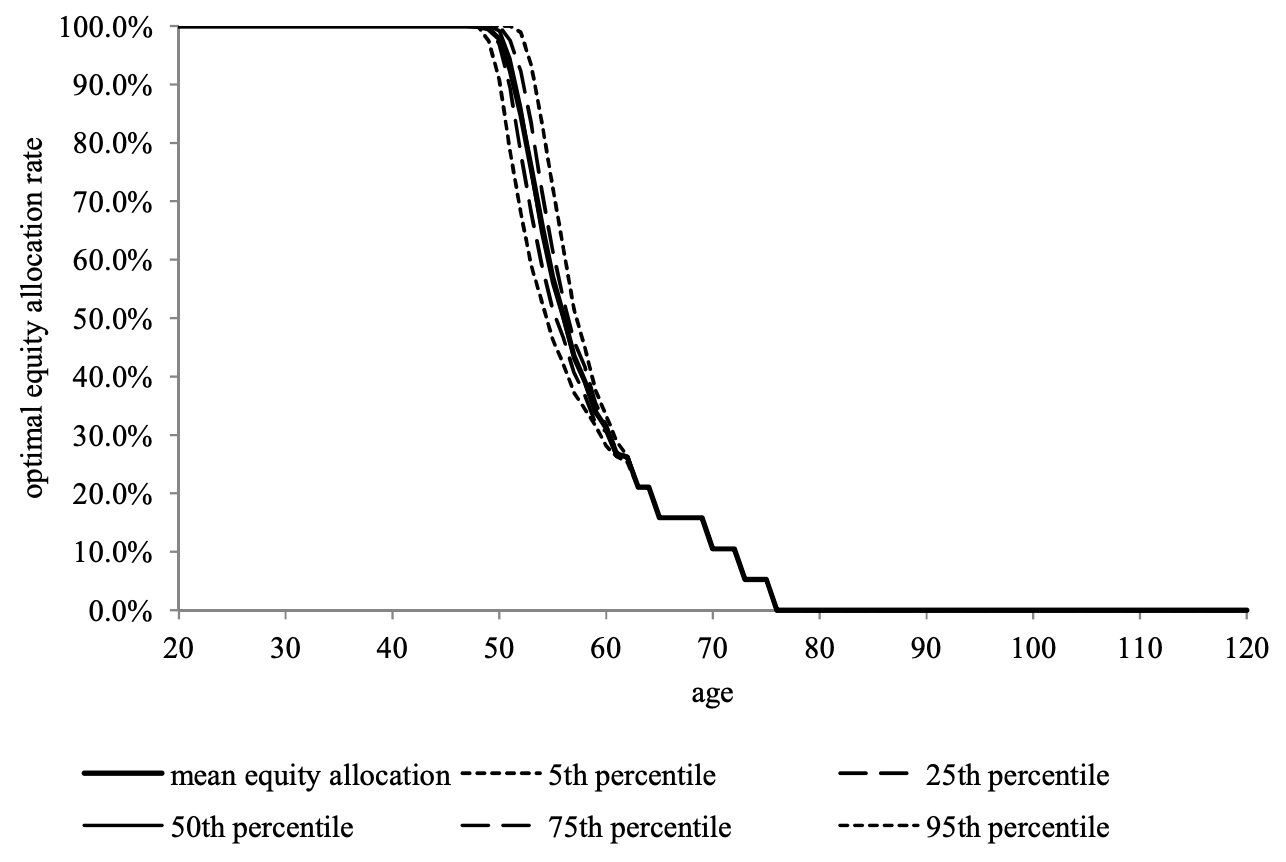}
    \caption{Optimal equity allocation rate over the life cycle (mean and percentiles from 10,000 simulations) according to \cite{blake}. The results illustrate a stochastic lifestyling strategy, with a high initial equity allocation decreasing significantly after age 45. }
    \label{fig:fig5_blake}
\end{figure}

Additionally, \cite{blake} present a model for optimal pension planning that highlights human capital as a key asset. Their study examines optimal funding and investment strategies within defined contribution pension plans by modeling participants as rational lifecycle financial planners who account for their human capital. Employing an Epstein–Zin utility function, the model conceptualizes human capital as the net present value of future labor income, which exhibits bond-like characteristics for younger individuals due to the relative stability of expected earnings. Consequently, the model demonstrates age-dependent optimal strategies: investment allocation initially favors equities to counterbalance the implicit bond exposure represented by human capital, gradually shifting toward bonds as human capital declines over the working life (see Figure~\ref{fig:fig5_blake}). 
Furthermore, the theoretical optimal funding strategy identified in \cite{blake} is also age-dependent, suggesting that contribution rates should ideally increase as retirement approaches—emphasizing the dynamic interaction between accumulating financial wealth and diminishing human capital throughout the lifecycle. The authors acknowledge that this contrasts with the common real-world practice of fixed-percentage contribution rates. Recognizing potential behavioral difficulties for individuals in implementing the purely optimal, heavily back-loaded contribution schedule, the paper suggests a practical compromise: a compulsory minimum contribution rate at all ages combined with the possibility of age-related additional voluntary contributions in later working years.

Despite significant theoretical advancements, the numerical implementation of these models remains challenging due to the complexity of dynamic programming techniques. The so-called “curse of dimensionality” makes solving models with many state variables or constraints computationally impractical. To address this, \cite{calafiore1} introduced an affine parameterization of the recourse policy, transforming the problem into a convex quadratic program and improving tractability. This approach simplifies the model by employing linear control policies, enabling efficient solutions involving many securities and time periods. \cite{calafiore2} later extended the model by incorporating transaction costs and relaxing the assumption of serially independent returns, allowing it to capture intertemporal market dependencies while maintaining computational efficiency.

Building on these foundations, several authors have explored more sophisticated dynamic asset allocation strategies. For example, \cite{kim} investigated how the investment horizon influences portfolio allocations, particularly in the context of mean-reverting stochastic risk premiums. Their study demonstrated that long-horizon investors—such as those saving for retirement—may hold different portfolios than short-horizon investors. Similarly, \cite{pfau} proposed valuation-based asset allocation strategies, showing that adjusting portfolio allocations based on market valuations (e.g., using cyclically adjusted price-to-earnings ratios) can improve retirement outcomes. \cite{yoon} developed a risk-budgeting approach for target-date funds, dynamically rebalancing portfolios to maintain predefined risk budgets. This strategy provided better downside protection during the 2008 financial crisis compared to conventional target-date funds.

\subsection{The Current Landscape of the Target-Date Fund Industry and Its Critiques}

Target-date funds (TDFs) have become a cornerstone of retirement planning, evolving from simple theoretical models into widely adopted investment products. As mentioned above, they were introduced by \cite{patent} in the late 1990s. TDFs aim to simplify retirement investing by automatically adjusting asset allocation based on a target retirement date.

As noted in the introduction, TDF adoption accelerated after the Pension Protection Act of 2006 designated them as a Qualified Default Investment Alternative in U.S. 401(k) plans, lowering employer barriers to offering them \citep{pensionact}. Prior to the Act, life-cycle funds had remained a niche product, struggling to gain broader market acceptance.

Early versions followed basic strategies, most notably the “100 minus age” rule, in which equity exposure declined as the investor aged. This rule suggests allocating a percentage of one’s portfolio to equities equal to 100 minus the investor’s age: for example, a 30-year-old would hold 70\% in stocks and 30\% in bonds. In the years that followed, researchers such as \cite{shiller} examined the performance of life-cycle portfolios in the context of proposed Social Security reforms. His findings highlighted the limitations of the 100-minus-age model, with returns often falling short of the offset rates needed to ensure adequate retirement income. The analysis underscored the challenges of implementing a standardized investment approach capable of serving a diverse workforce under varying economic conditions.

Over time, the industry saw substantial changes. By 2012, TDFs accounted for 20\% of all 401(k) assets, with 72\% of plans offering them. These funds evolved beyond simple equity-bond allocations to incorporate a broader mix of asset classes, including alternatives like real estate and commodities. TDFs offered certain advantages, including access to institutional share classes with lower expense ratios, but they often underperformed passive index strategies with equivalent allocations. Furthermore, some investment decisions within TDFs, particularly regarding fund families, were sometimes seen as benefiting the fund managers rather than the investors \citep{elton}.

The industry’s growth was also marked by increasing diversity in risk profiles and investment strategies. As noted by \cite{balduzzi}, there was significant heterogeneity among TDFs with the same target date, especially after 2006. Their research revealed that newer fund families with smaller market shares often took on more idiosyncratic risks, leading to varying returns within the same target-date cohort. This variability underscored the need for greater transparency regarding the risk profiles of TDFs. Similarly, \cite{estrada2} found that the declining-equity glide path commonly used in TDFs aligned with investors' increasing risk aversion as they aged, but raised questions about whether this approach was truly optimal given the evolving nature of risk tolerance over the course of retirement.

The growth of TDFs was further evidenced by their size, reaching \$1.5 trillion by 2020, up from \$0.3 trillion in 2011 in the U.S., and by 2022, according to Vanguard, 95\% of pension sponsors offered TDFs, and 98\% of investors selected a TDF as their default option \citep{lessons}.  These funds became a cornerstone of retirement planning, with managers appearing to have learned from previous market crises. Notably, during the COVID-19 pandemic, TDF managers generally adhered to their glide paths, resisting the temptation to chase returns. However, despite these improvements, \cite{lessons} observed significant variability in performance among funds with the same target date, particularly following the market downturn in March 2020. While the funds recovered quickly, underscoring the resilience of TDFs when managed according to their glide paths, the findings also highlighted the diversity in risk-adjusted outcomes.

\subsubsection{Critiques of Current Glide Path Designs}

Despite the success and growth of TDFs, a surge of criticism has emerged, particularly concerning the construction of glide paths. Traditional glide paths, which reduce equity exposure as investors age, have been questioned for their failure to adapt to an evolving financial landscape. For example, \cite{pfau2} challenges conventional wisdom by suggesting that a rising equity glide path during retirement can outperform traditional strategies. By starting with a conservative equity allocation and gradually increasing exposure, retirees may reduce the risk of portfolio failure, especially in the early years of retirement, while still benefiting from growth in later years.

Similarly, \cite{estrada3} provides evidence that contrarian strategies, such as those that increase, rather than decrease, equity exposure over time, tend to deliver better terminal wealth outcomes than traditional life cycle strategies. His research, covering 19 countries over a 110-year period, demonstrated that while such strategies involve higher volatility, they generally offer a more favorable risk–return profile, particularly during periods of market growth. These findings raise important questions about whether the widely adopted glide path approach is truly the optimal strategy for retirement planning.

Additionally, \cite{estrada1} found that simpler strategies, such as static 60/40 portfolios or all-equity portfolios, often outperform more complex glide paths. This challenges the notion that multi-faceted glide path strategies are necessary for optimal retirement outcomes. These findings suggest that a simpler, more straightforward approach may better serve investors, particularly in terms of performance consistency over time.

However, there are also voices in the industry advocating for greater complexity in glide paths. \cite{forsyth} argues that traditional allocation strategies, such as the 60/40 portfolio or fixed glide paths, fail to adapt to changing market conditions and accumulated wealth. His research, which incorporates both parametric modeling and bootstrap resampling, found that adaptive strategies—those that adjust based on both time and current portfolio value—outperformed traditional approaches by reducing the standard deviation of terminal wealth. Forsyth’s conclusion that most TDFs fail to adequately serve investors raises concerns about the fundamental design of these funds and their effectiveness as default retirement vehicles.

All in all, target-date funds have become a critical component of modern retirement plans, but their construction, particularly the glide paths, remains a contentious and controversial topic. Some critics argue for more flexibility and adaptive strategies, whereas other authors support more simplified approaches, both questioning the widespread adoption of the current glide path model. As the industry continues to evolve, the debate over the optimal construction of target-date funds is likely to persist.

\section{Target-Date Funds in the Context of Latin American Pension Systems}

In the last years, several Latin American countries—Mexico, Costa Rica, Colombia, Peru, Uruguay, and Chile—have adopted defined contribution (DC) pension systems with investment structures that seek to align portfolio risk with the age of participants \citep{mantilla}. While most of these arrangements are often compared to target-date funds, they are more accurately characterized as multifund systems: each fund maintains a static asset allocation, and individuals are periodically reallocated between funds of varying risk levels as they approach retirement. In contrast, true life-cycle strategies involve a continuous adjustment of the portfolio’s risk profile within a single fund. Among the countries listed, only Mexico has implemented such a design, through its generational SIEFORES. However, Chile —historically structured as a multifund system— is currently undergoing a regulatory transition toward a target-date framework. Given this distinction, now we will focus on the investment regimes of Mexico and Chile, which offer the most relevant cases for understanding life-cycle investing in Latin American pension systems.

\subsection{Investment Structure of the Mexican Pension System}

The current investment regime of the Mexican pension system is structured around individualized retirement accounts, which are privately managed by Administradoras de Fondos para el Retiro (AFOREs). Each AFORE invests these resources through diversified vehicles known as Sociedades de Inversión Especializadas en Fondos para el Retiro (SIEFOREs). Historically, SIEFOREs were classified by age segments, but since December 2019, Mexico has transitioned to a generational fund (or target-date fund) model, in which each SIEFORE functions as a TDF. Under this new structure, workers are automatically assigned to funds based on their birth year, dynamically aligning portfolio risk profiles with their investment horizons.

In this context, each SIEFORE is designed with a specific glide path that gradually shifts its asset allocation from riskier to more conservative investments as the target retirement date approaches. At inception, younger participants’ funds have higher allocations to equities, foreign assets, and structured products, aiming to maximize long-term returns. Conversely, as participants near retirement age, fund portfolios increasingly shift toward safer instruments such as government bonds and inflation-linked securities, in order to protect accrued capital and minimize risk exposure.

There are 10 SIEFOREs in total. The youngest participants are placed in the Initial Basic SIEFORE, which, as of 2025, corresponds to individuals born in the year 2000 or later. From there, the remaining SIEFOREs are organized into five-year generational brackets. For example, in 2025, the subsequent brackets are: 1995–1999, 1990–1994, 1985–1989, 1980–1984, and so on.

Mexico’s regulated pension system manages investment risk through a standardized set of regulatory limits for each SIEFORE, as defined by the National Retirement Savings System Commission or CONSAR \citep{consar}. These limits—covering asset class exposures and risk metrics such as Value-at-Risk (VaR) and CVaR Differential—are adjusted quarterly to ensure a gradual reduction in portfolio risk as participants age, with the adjustment schedule known in advance. Each generational SIEFORE enters this regulatory trajectory at a distinct starting point based on the participant’s birth year, meaning the applicable limits for any given SIEFORE are determined by both its generational classification and its current position within the quarterly glide path. In this way, the regulator enforces a unified limit path through these evolving investment constraints. To ensure further compliance, CONSAR mandates that each SIEFORE portfolio maintain a maximum tracking error of 5\% relative to the \textit{Investment Trajectories} (internal benchmarks) established by each AFORE’s governing body. Tables~\ref{tab:investment_limits_1} and~\ref{tab:investment_limits_2} provide a snapshot of the current regulatory limits across major asset classes for different generational cohorts.

\begin{table}[htbp]
\centering
\caption{Limits by Type of Basic SIEFORE: From Initial to 80–84 Age Bracket}
\label{tab:investment_limits_1}
\begingroup
\renewcommand{\arraystretch}{1.3}
\footnotesize
\begin{tabular}{>{\bfseries}l|c|c|c|c|c}
SIEFORE & Initial & 95–99 & 90–94 & 85–89 & 80–84 \\
\hline
Quarter & 1 & 2 & 22 & 42 & 62 \\
\hline
\multicolumn{6}{l}{\textbf{Risk Limits}} \\
\hline
Value at Risk& \multicolumn{5}{c}{--} \\
Diff. in CVar& 
1.3\% & 1.3\% & 1.3\% & 1.2\% & 1.2\% \\
\hline
\multicolumn{6}{l}{\textbf{Asset Class Limits}} \\
\hline
Foreign Securities& \multicolumn{5}{c}{20\%} \\
Equity& 
60.0\% & 60.0\% & 58.2\% & 56.0\% & 53.1\% \\
Foreign FX& 
35.0\% & 35.0\% & 34.4\% & 33.7\% & 32.8\% \\
Securitizations& 
40.0\% & 40.0\% & 38.3\% & 35.8\% & 33.0\% \\
Structured& 
20.0\% & 20.0\% & 20.0\% & 20.0\% & 20.0\% \\
FIBRAS& 
10.0\% & 10.0\% & 10.0\% & 10.0\% & 10.0\% \\
Inflation Protected& \multicolumn{5}{c}{--} \\
Commodities& \multicolumn{5}{c}{5\%} \\
\end{tabular}
\endgroup
\end{table}

\begin{table}[htbp]
\centering
\caption{Limits by Type of Basic SIEFORE: From 75-79 to Pension Bracket}
\label{tab:investment_limits_2}
\begingroup
\renewcommand{\arraystretch}{1.3}
\footnotesize
\begin{tabular}{>{\bfseries}l|c|c|c|c|c}
SIEFORE & 75–79 & 70–74 & 65–69 & 60–64 & Pension \\
\hline
Quarter & 1 & 2 & 22 & 42 & 62 \\
\hline
\multicolumn{6}{l}{\textbf{Risk Limits}} \\
\hline
Value at Risk&  \multicolumn{3}{c}{--}  & 0.9\% & 0.7\% \\
Diff. in CVar& 
1.0\% & 0.8\% & 0.5\% & 0.2\% & 0.1\% \\
\hline
\multicolumn{6}{l}{\textbf{Asset Class Limits}} \\
\hline
Foreign Securities& \multicolumn{5}{c}{20\%} \\
Equity& 
49.4\% & 43.3\% & 34.2\% & 18.9\% & 15.0\% \\
Foreign FX.& 
31.6\% & 28.6\% & 24.3\% & 17.3\% & 15.0\% \\
Securitizations& 
30.2\% & 26.7\% & 23.1\% & 20.7\% & 20.0\% \\
Structured& 
20.0\% & 17.1\% & 14.3\% & 11.4\% & 10.0\% \\
FIBRAS& 
10.0\% & 8.6\% & 7.1\% & 5.7\% & 5.0\% \\
Inflation Protected& \multicolumn{3}{c|}{--} & \multicolumn{2}{c}{Min 51\%} \\
Commodities& \multicolumn{5}{c}{5\%} \\
\end{tabular}
\endgroup
\end{table}

The investment limits for SIEFOREs have evolved significantly to allow for greater diversification. However, despite regulatory advances, the portfolios of Mexican pension funds remain notably conservative. For example, current limits still appear strict compared to TDFs in the United States: in Mexico, the equity cap for the Initial SIEFORE is 60\%, whereas the typical equity allocation in the early stages of U.S. TDFs ranges from 85\% to 95\%. Government debt continues to dominate, accounting for over 50\% of total assets under management—significantly higher than the OECD average. Moreover, foreign investment remains capped at 20\%, limiting international diversification and potential returns \citep{bid2019diagnostico}.

Operationally, the generational funds approach faces certain frictions. For instance, fund transfers between generations—automatically triggered as participants age—can generate transaction costs and short-term market risks. Furthermore, the risk measures currently employed by CONSAR (e.g., Value-at-Risk and Conditional Value-at-Risk) primarily focus on short-term losses rather than long-term retirement adequacy. As a result, a gap remains between the regulatory investment framework and the ultimate pension outcomes desired by contributors.

\subsection{The Chilean Pension System and Its Reforms}

Before 1981, Chile's pension system was fragmented, with around 35 pension funds and over 150 distinct regimes. These systems, funded by mandatory contributions from both employees and employers, provided defined benefits for pensions and insurance, but were criticized for being complex and inequitable. By 1980, the system covered 75\% of workers but faced fiscal and administrative challenges. In 1981, Chile implemented a major reform, replacing the pay-as-you-go model with a privately managed, individual capitalization system. Workers contribute 10\% of their income to individual accounts managed by private pension fund managers (AFPs), with fees initially as high as 4.87\%, later stabilizing at around 2.5\% \citep{bravo2015informe}.

In 2002, Chile introduced a multifund system, offering five investment portfolios (A, B, C, D, and E) with varying asset allocations and risk levels to enable more personalized retirement planning. The 2008 reform added a solidarity pillar by establishing the Basic Solidarity Pension and the Solidarity Pension Contribution, both funded by general taxation, which improved coverage for low-income groups. This reform also introduced mandatory employer-funded disability and survivor insurance.


Following the introduction of multifunds in 2002, Chile established a default fund assignment strategy based on age and gender for individuals who do not actively select a specific pension fund. This multifund glide path aims to balance risk and return across different life stages. The default fund allocation is explained in Figure~\ref{fig:multifund-glidepath}.

\usetikzlibrary{positioning}

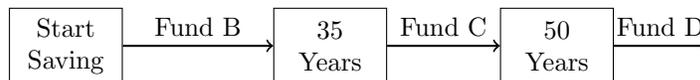
\begin{figure}[h!]
\centering
\begin{tikzpicture}[node distance=1.5cm]
    \node (titleMen) at (3, 1.8) {\textbf{Glide Path for Men}};
    \node (startMen) [draw, rectangle, minimum height=1cm, minimum width=1.5cm, align=center] {Start\\Saving};
    \node (middleMen) [right=2cm of startMen, draw, rectangle, minimum height=1cm, minimum width=1.5cm, align=center] {35\\Years};
    \node (endMen) [right=2cm of middleMen, draw, rectangle, minimum height=1cm, minimum width=1.5cm, align=center] {55\\Years};
    
    \draw[->, thick] (startMen) -- (middleMen) node[midway, above] {Fund B};
    \draw[->, thick] (middleMen) -- (endMen) node[midway, above] {Fund C};
    \draw[->, thick] (endMen) -- ++(2,0) node[midway, above] {Fund D}; 
    
    \node (titleWomen) at (3, -1.5) {\textbf{Glide Path for Women}};
    \node (startWomen) [below=2cm of startMen, draw, rectangle, minimum height=1cm, minimum width=1.5cm, align=center] {Start\\Saving};
    \node (middleWomen) [right=2cm of startWomen, draw, rectangle, minimum height=1cm, minimum width=1.5cm, align=center] {35\\Years};
    \node (endWomen) [right=1.5cm of middleWomen, draw, rectangle, minimum height=1cm, minimum width=1.5cm, align=center] {50\\Years};
    
    \draw[->, thick] (startWomen) -- (middleWomen) node[midway, above] {Fund B};
    \draw[->, thick] (middleWomen) -- (endWomen) node[midway, above] {Fund C};
    \draw[->, thick] (endWomen) -- ++(2,0) node[midway, above] {Fund D}; 
\end{tikzpicture}
\caption{Default Glide Path Allocation for Women and Men Based on Age}
\label{fig:multifund-glidepath}
\end{figure}

This simple approach implicitly transitions savings from higher-risk portfolios to more conservative investments as individuals approach retirement age, thus providing a structured pathway towards reducing investment risk over time. While not strictly a target-date structure by definition, it effectively creates a glide path, gradually adjusting the asset allocation to align with the changing risk profile of the investor as they move closer to retirement.

In 2022, the Universal Guaranteed Pension (PGU) replaced the previous solidarity system, extending benefits to nearly all Chileans aged 65 or older, regardless of employment status. Funded by the state, the PGU aims to significantly expand coverage for the elderly population and ensure broader social protection \citep{ley2022reforma}.

In 2025, Chile approved a significant pension reform, which will be explained in the following section.

\subsection{The 2025 Chilean Pension Reform}

In January 2025, Chile implemented a significant pension reform aimed at addressing long-standing challenges in investment management, gender equity, and pension adequacy. This reform includes a series of changes to contribution structures and introduces a new investment framework, including measures to enhance AFP operations \citep{ley2025pensiones}.

According to the pension reform, a fundamental point to emphasize is the long-term increase in the total mandatory savings contribution rate directed to individual accounts, rising from the current 10\% to an aggregate 16\% of income. This significant increase occurs as the existing 10\% contribution will be supplemented by an additional 6\% mandatory savings contribution funded by the employer. Although the total new mandatory contribution is set at 18.5\%, 2.5\% will be allocated to the Social Insurance fund. The remaining 6\% consists of a 4.5\% direct employer contribution to the individual account and a 1.5\% transitional deferred contribution known as the “Contribution with Protected Return” (\textit{Aporte con Rentabilidad Protegida}), which is also ultimately destined for individual accounts. This 1.5\% contribution is transitional for 30 years and will be disbursed upon the worker’s retirement to fund a benefit based on the number of years contributed. Consequently, the most significant long-term implication for individual retirement savings is the elevation of the total mandatory contribution rate to 16\%.

To support vulnerable groups, the \textit{Pensión Garantizada Universal (PGU)}, a state-funded, non-contributory pension aimed at ensuring a minimum income for elderly individuals, will gradually increase to CLP \$250,000. In addition to the higher benefit amount, the reform expands the eligibility criteria, allowing broader coverage of vulnerable populations who were previously excluded due to income or other requirements.

A major innovation of the reform is the shift from multifunds to target-date funds—namely, \textit{generational funds}—in which individuals are automatically assigned to a fund based on their birth year. This glide path strategy ensures that investment risk is progressively adjusted as participants approach retirement, offering a more appropriate risk profile throughout their lives. These generational funds are designed to reduce the need for frequent fund switches, providing a stable and structured investment pathway for contributors.

The definition of the glide path will be incorporated into a benchmark outlined in the new Investment Regime, which will be established by the Superintendencia de Pensiones in 2026. This regulatory body will determine the exact parameters of the glide path, including the asset allocation and risk adjustment over the participants' life cycle. It must also determine the maximum allowable deviations from that benchmark. The implementation timeline includes key milestones from May 2025 to August 2027, with gradual increases in employer contributions, the introduction of new funds, and scheduled payments for those affected by the reform.

In addition to structural changes, the reform modifies the reserve (known as the ``\textit{encaje}'') mechanism that AFPs must maintain. Instead of a fixed percentage of each fund’s assets, the total reserve now corresponds to 30\% of the AFP’s commissions collected over the previous 12 months. This total is then proportionally allocated across all funds managed by the AFP, based on their relative size. According to the reform, the purpose of this reserve is to cover any mandatory contributions that the AFP must make to a fund when its monthly return falls below a minimum “performance unit” defined in the Investment Regime. In this way, the reform links fund underperformance to AFP accountability through this reserve mechanism.

\section{Considerations for Designing the Glide Path and Investment Regime in the Chilean Context}

The design of the glide path and Investment Regime for Chile’s target-date funds requires careful attention to both theoretical foundations and practical implementation. Several key considerations should guide the regulator in defining the optimal glide path for participants. First and foremost, the glide path must align with the evolving risk profiles of individuals as they approach retirement. Younger participants—who can tolerate higher levels of risk and possess greater potential in terms of human capital—should have a larger allocation to equities, while older participants require more conservative allocations to preserve capital. However, the regulator should critically assess whether the widely used approach of excessively slow declines in equity exposure is indeed the most effective. As mentioned earlier, recent studies suggest that alternative glide paths—such as those maintaining or even increasing equity exposure toward retirement—may deliver superior outcomes under certain market conditions \citep{estrada3}.

It is also important to assess what level of equity allocation is appropriate, and whether assigning a relatively low percentage to equities for the youngest participants—such as in the case of Mexico—is excessively conservative, potentially falling short of generating sufficient returns for future pensions, especially when there is no clear justification for such a low maximum equity level (60\% in Mexico’s case).

Additionally, it is essential to incorporate dynamic risk management strategies into the design of both the glide path and investment limits. This could include integrating risk measures such as Conditional Value-at-Risk (CVaR), which more effectively capture potential market volatility. By adjusting the glide path in response to changing market conditions, funds can maintain their intended risk-return profile. Moreover, the glide path should be flexible enough to accommodate shifts in the broader economic environment—such as inflation, market downturns, or other external shocks—ensuring that the funds continue to meet participants' retirement goals.

Another crucial consideration is the need for adequate diversification—not only across domestic asset classes but also through international exposure and the inclusion of alternative assets. International equities and foreign debt instruments offer significant diversification benefits, enhancing returns while mitigating risks associated with over-reliance on local markets. Furthermore, incorporating alternative assets such as real estate, infrastructure, private debt, and private equity could improve portfolio performance by providing additional sources of return and reducing correlation with traditional asset classes. However, the integration and benchmarking of alternative assets require careful consideration: indeed, empirical evidence from pension funds suggests that selecting appropriate benchmarks for private equity is particularly challenging in practice, influenced by factors beyond simple performance replication, such as advisor incentives and varying selection processes \citep{augustin2023benchmarking}. Given these complexities and the inherent difficulty of replicating the performance of these assets, it is advisable to allow for broader performance bands within the funds’ benchmarks. This flexibility would enable funds to implement their alternative asset strategies more effectively, enhancing their ability to adapt to evolving market conditions while maintaining the intended risk-return profile.

Lastly, research suggests that portfolio adjustments based on an individual’s wealth can optimize retirement outcomes, as they tailor the risk profile to the investor’s current financial situation \citep{cocco}. Such an approach allows for more personalized risk management: participants with greater accumulated wealth may afford to take on more risk, while those with less may need to prioritize capital preservation. By accounting for wealth accumulation, the glide path can better align with the varying needs of participants at different stages of their retirement savings journey. While it may not be feasible to personalize the glide path for each participant in the Chilean context, it is important to consider that different pension fund managers (AFPs) could adopt distinct glide paths based on the wealth characteristics of their client base—thereby allowing for greater differentiation among managers.

Considering all the points discussed in this section, the regulator should carefully define investment limits, asset allocation strategies, and glide path adjustments to ensure alignment with long-term retirement adequacy goals. The implementation of these parameters should be transparent and evidence-based, drawing on both international best practices and the specific needs of Chilean workers. Such a balanced approach will help ensure that generational funds deliver optimal retirement outcomes for participants across a range of economic conditions.

\section{Conclusion}

The 2025 Chilean pension reform, with its introduction of target-date funds (TDFs), represents a significant advancement in the management of retirement savings. To ensure the success of this transition, several key recommendations should be considered.

First, the glide path design should be dynamic, allowing for adjustments based on market conditions. While the traditional approach of gradually reducing equity exposure with age is common, emerging research suggests that alternative equity exposure frameworks may lead to improved retirement outcomes. A more flexible glide path—one that maintains higher exposure to risky assets for a longer period—would better align with the evolving financial situations of participants and broader market dynamics.

Second, the regulator should consider adopting a wide performance deviation band for the benchmark of each generational fund. This would foster competition among fund managers, enabling them to differentiate themselves based on performance, asset allocation strategies, and value propositions tailored to the characteristics of their client base—such as individual wealth levels. This flexibility would enhance diversification, improve potential returns, and reduce over-reliance on the benchmark definition.

Finally, future work should focus on defining a glide path model that incorporates state-of-the-art theoretical foundations—such as those discussed in this paper—while also reflecting the specific parameters of the Chilean pension system. Such a model would integrate dynamic asset allocation strategies, modern risk management techniques and measures, and a broader range of asset classes, ensuring optimal alignment with the evolving needs of Chilean participants. The successful implementation of target-date funds under the 2025 reform will ultimately depend on a flexible glide path, diversified asset allocations, and the ability of fund managers to innovate. These elements—alongside the development of a comprehensive, context-sensitive glide path model—could significantly enhance the Chilean pension system and improve long-term retirement outcomes for its workers.

\bibliographystyle{elsarticle-harv}
\bibliography{
bib/samuelson1,
bib/MERTON1,
bib/patent,
bib/pensionact, 
bib/markowitz, 
bib/lessons,
bib/reforma,
bib/consar,
bib/estrada1,
bib/elton,
bib/ramsey,
bib/merton2,
bib/gomes,
bib/cocco,
bib/calafiore1,
bib/calafiore2,
bib/kim,
bib/pfau,
bib/yoon,
bib/manu,
bib/shiller,
bib/balduzzi,
bib/estrada2,
bib/blake,
bib/pfau2,
bib/estrada3,
bib/forsyth,
bib/ley2022reforma,
bib/ley2025pensiones,
bib/bravo2015informe,
bib/bid2019diagnostico,
bib/mantilla
}

\end{document}